\documentclass[aps,prd,twocolumn,preprintnumbers,superscriptaddress]{revtex4}
\usepackage{graphicx}
\usepackage{subfigure}
\usepackage{epsfig}
\usepackage{amssymb}
\usepackage{amsmath}
                                                                                
\begin{document}
\preprint{IGPG-05/7-1}
\preprint{AEI-2005-122}

\newcommand{\lP}{l_{\mathrm P}}
\newcommand{\lp}{l_{\mathrm P}}                                                                 
\newcommand{\md}{{\mathrm{d}}}
\newcommand{\tr}{\mbox{tr}}
\newcommand{\be}{\begin{equation}}
\newcommand{\ee}{\end{equation}}
\newcommand{\bq}{\begin{eqnarray}}
\newcommand{\eq}{\end{eqnarray}}
\newcommand{\phieff}{\phi_{{\mathrm{eff}}}}
                                                                                
\newcommand*{\R}{{\mathbb R}}
\newcommand*{\N}{{\mathbb N}}
\newcommand*{\Z}{{\mathbb Z}}
\newcommand*{\Q}{{\mathbb Q}}
\newcommand*{\C}{{\mathbb C}}
                                                                                
\newcommand{\ket}[1]{| #1 \rangle}
\newcommand{\bra}[1]{\left\langle #1 \right|} \newcommand{\braket}[2]{\langle #1 \,|\, #2 \rangle }
\newcommand{\braketfull}[3]{\langle #1 \,|\, #2 \,|\, #3 \rangle }
                                                                                
\newcommand{\bas}{\ket{\mu}}
\newcommand{\half}{\frac{1}{2}}
\newcommand{\sgn}{\text{sgn}}
\newcommand{\cof}[1]{s_1(\mu #1) \, \psi_{\mu #1}}
\newcommand{\cs}{\widehat{\text{cs}}}
\newcommand{\sn}{\widehat{\text{sn}}}
\newcommand{\Pp}{P_{\phi}}
                                                                                
\title{Semi-classical States, Effective Dynamics and Classical Emergence in Loop Quantum Cosmology}
\author{Parampreet Singh}\email{singh@gravity.psu.edu}
\affiliation{Institute for Gravitational Physics and Geometry, The Pennsylvania
State University, 104 Davey Lab, University Park, PA 16802, USA}
\author{Kevin Vandersloot}\email{kfvander@gravity.psu.edu}
\affiliation{Institute for Gravitational Physics and Geometry, The Pennsylvania
State University, 104 Davey Lab, University Park, PA 16802, USA}
\affiliation{Max-Planck-Institut f\"ur Gravitationsphysik,
Albert-Einstein-Institut,\\
Am M\"uhlenberg 1, D-14476 Golm, Germany}

\begin{abstract}
We construct physical semi-classical states annihilated by the
Hamiltonian constraint operator in the framework
of loop quantum cosmology as a method of systematically
determining the regime and validity of the semi-classical
limit of the quantum theory. 
Our results indicate that the evolution can be
effectively described using continuous classical equations of motion
with non-perturbative corrections down
 to near the Planck scale 
below which the universe can only be described by the discrete quantum constraint.
These results, for the first time, provide  concrete evidence of the emergence
of classicality in loop quantum cosmology and also
clearly demarcate the domain of validity of different effective 
theories. We prove the validity of modified Friedmann dynamics incorporating
discrete quantum geometry effects which can lead to various new 
phenomenological applications.
Furthermore the understanding of semi-classical states
allows for a framework for interpreting the quantum wavefunctions
and  understanding questions of a semi-classical
nature within the quantum theory of loop quantum cosmology.

\end{abstract}

\maketitle

\section{Introduction}

A fundamental input of loop quantum gravity (LQG) 
\cite{lqg_review} to our understanding of
quantum spacetime is that it is inherently discrete 
and spatial geometry is quantized.
Quantum  features of spacetime become evident in the 
regime of very high curvature
whereas continuous spacetime emerges as a large eigenvalue 
limit of quantum geometry. 
Perhaps the most interesting avenue to explore this idea 
is in cosmological FRW spacetimes in the regime of high curvature
and small volume 
such that quantum gravitational effects are expected to be dominant
with the possibility of potentially observable signatures.
It is a fundamental question in any viable model of LQG
as to the scale at which the classical picture can be
recovered and where precisely do we expect to see
modifications to the classical Friedmann dynamics. 
Since tools to address these 
issues are still under development \cite{thomas}, it
is an open question whether the picture suggested by LQG 
holds for our Universe.

To construct a FRW model within the field theoretic framework
of LQG would be quite difficult. Thus much progress has
been made by restricting the model to a mini-superspace
quantization known as loop quantum 
cosmology (LQC). In this simplified setting fundamental
questions can be answered directly with explicit calculations
and the physical consequences can be explored. 
As in LQG, the underlying geometry in LQC 
is discrete and the scale factor operator has discrete 
eigenvalues. Quantum dynamics is governed by a discrete 
difference equation which leads to a non-singular evolution 
through the classical big bang 
singularity \cite{bigbang,Bohr}. This important result
can be traced directly to the discrete nature of quantum geometry. 
A related and important feature
of LQC is the modification to the behavior of the eigenvalues of the
inverse scale factor below a critical scale factor
$a_*$. Unlike in the classical regime, eigenvalues of the inverse 
scale factor operator become 
proportional to positive powers of scale factor for $a < a_*$.
These considerations have led to an effective description
of the evolution of the Universe, where the standard
Friedmann dynamics receives modifications from the inverse
scale factor eigenvalues below $ a_*$. Hence it has been
assumed that classical emergence occurs above
$ a_*$ and that the universe can still
be described in terms of continuum dynamics
below this scale.

The effective modified Friedmann dynamics leads to various 
interesting phenomenological effects (see 
Ref.\cite{martin_review} for a recent review). For example, it 
has been demonstrated that effective dynamics
leads to a phase of kinetic dominated super-inflation
\cite{superinflation} 
which can provide correct initial
conditions for conventional chaotic inflation 
\cite{closedinflation,inflationcmb,Robust,effham1}, 
thus making inflation
more natural.  
The loop quantum phase has been shown to 
alleviate the problem of inflation in
closed models \cite{infbounce1} and inflation without scalar field \cite{density}.
It is interesting to note that 
effects of a super-inflationary phase
prior to the conventional potential driven inflation may leave 
observable signatures at large scales in the cosmic 
microwave background \cite{inflationcmb}. Apart from effects in 
the early Universe, the effective dynamics promises
to resolve various cosmological singularities, for example the big crunch 
\cite{BounceClosed,BounceQualitative,infbounce2,effham2} and
brane collision singularity \cite{Cyclic}. Application of LQC 
techniques have also shown to yield non-singular
 gravitational collapse scenarios
with associated observable
signatures \cite{bhole,naked}.

Although LQC phenomenology leads to various potentially observable 
effects, it should be
noted that strictly speaking these investigations are 
not based directly on the quantum theory of LQC, but
rather on  heuristically motivated effective continuous
equations of motion. Beyond heuristic ideas it is not clear
as to the domain of validity of the effective theory and
where the continuous effective description breaks down. Thus
it merits a more careful derivation of the effective
picture directly from the quantum theory and
difference equation. In the absence of any such 
quantitative proof, the above assumptions on the 
existence of the modified dynamics below $a_*$
appear ad-hoc. These issues have been noted before and in fact phenomenological constraints 
on the domain of validity have also been discussed \cite{Robust,kevin_ham},
but in the absence of any comparison with the underlying
quantum evolution, such investigations are far from being
complete.

A further issue less emphasized previously pertains to the fact
that LQC predicts a discrete difference equation as opposed
to the continuous Wheeler-DeWitt equation of standard quantum cosmology.
It is thus important to determine under what conditions can the
discreteness modify the dynamics and play an important role
phenomenologically. An important question is whether the continuum picture
breaks down in the regime where discreteness effects are important and
if not how can  the dynamics  be described in terms of an effective
continuous picture.

The aim of this paper is to answer these questions
systematically through the application of the full quantum
features of LQC. To this end we will study the model
of a homogeneous and isotropic universe with matter
coupled in the form of a massless scalar field. The main
difficulty in describing dynamics within LQC pertains
to oft mentioned problem of time in quantum cosmology
\cite{probtime}. The problem can however be overcome by
treating the scale factor as clock variable and considering
the evolution of the scalar field. Thus the semi-classical
states constructed will consist of sharply peaked wave packets
centered around a value of the scalar field at a particular scale
factor. The
states can be evolved forward (or backward) using the difference equation
and the trajectory can be compared with that from classical Friedmann or effective dynamics. All of the physics is captured in
this picture without resorting to an external time.
This allows us to directly verify the semi-classical
limit of LQC and determine where the continuum picture
breaks down. The results will validate the 
effective continuous picture incorporating the inverse volume modifications.
Furthermore, we will also test and verify previously proposed
effective continuous  equations that include modifications
associated with the discreteness effects. We will indicate 
with the quantum evolution  the conditions under which the discreteness
effects play an important role.


\section{Classical Theory and the Quantum Constraint}
The model we will investigate is composed of a homogeneous and 
isotropic universe
with zero spatial curvature and matter in the form of a massless
scalar field. The classical phase space is parameterized
by four quantities: the connection $c$, the triad $p$, 
the scalar field $\phi$, and the scalar field
momentum $\Pp$ which satisfy the following Poisson bracket relations
\be
\{ c, p\} = \frac{1}{3} \kappa \gamma, ~~ \{ \phi, \Pp \} = 1 \,.
\ee
Here $\gamma$ is 
the Barbero-Immirzi parameter whose
value can be fixed to $0.2375$  
by black hole thermodynamics \cite{bek_hawking} and
$\kappa = 8 \pi G$. We will work with $c = \hbar = 1$
and Planck length $\lp = \sqrt{G}$. Note that
formulas in LQC typically are written in terms
of the modified Planck length $\lp = \sqrt{8\pi G}$, hence
our formulas will appear slightly modified from
previous works.

The gravitational variables
$c$ and $p$ encode the curvature and geometry respectively
of the gravitational field which can be seen in their
relation to the standard metric variables
\be \label{pc}
|p| = a^2, ~~ c = \gamma \dot{a} \,.
\ee
Governing the dynamics is the Hamiltonian constraint 
which in terms of $c$ and $p$ variables is given by
\begin{equation}\label{Hclass}
        H = - \frac{3}{\kappa \gamma^2} \sgn(p) \sqrt{|p|} \; c^2
        + \half a^{-3} \Pp^2 \;.
\end{equation}
The equations of motion are derived through the vanishing
of the Hamiltonian constraint $H = 0$, and the Hamiltonian
equations ($\dot{x}=\{x,H\}$ for any phase space variable $x$).
For the scalar field momentum, the Hamiltonian equations
immediately imply that $\Pp$ is constant in time.
The Hamiltonian equations for $p$ and $\phi$ along
with the vanishing of the Hamiltonian constraint give
their time dependence
\be \label{clmotion}
\dot{p} = \pm\sqrt{\frac{2 \kappa}{3}} \frac{\Pp}{a},
~~ \dot{\phi} = a^{-3} \Pp
\ee
which can be integrated to give the time evolution
of both $p$ and $\phi$.
The standard Friedmann equation can be derived
by combining these equations, eliminating $\Pp$, 
and using the relation between $p$ and $a$ in
to get
\begin{eqnarray}
	\left( \frac{\dot{a}}{a} \right)^2 &=& \frac{\kappa}{6} 
	\,\dot{\phi}^2 \; ~.
\end{eqnarray}

Our goal is to compare the trajectories of quantum wave packets
in LQC with the classical equations of motion (\ref{clmotion}).
The difficulty lies in the lack of any $\partial/\partial t$
term in the difference equation that governs the behavior
of the wave packets. Thus it is impossible within LQC to
consider a wave packet peaked around some value of $p$ and
$\phi$ and evolve it forward in time to compare with the classical
trajectory. This is the ``problem of time" in quantum cosmology 
\cite{probtime}.
The origin of this difficulty can be understood even at the
classical level; namely that the classical equations of
motion are not unique since the lapse is a freely specifiable
function which we have implicitly fixed to one to arrive
at (\ref{clmotion}). The trajectory $p(t)$ itself has no
physical meaning since we can re-parameterize $t$ to get a
different trajectory. Physically an observer could never
measure the value of $t$ by measuring $p,c,\phi$, and $\Pp$.

A solution to the problem, as noted by various authors
\cite{qgtime}, is to notice that while
for instance $p(t)$ and $\phi(t)$ by themselves have no physical
meaning,
the correlation between $p$ and $\phi$ for a given value of $t$ is
a physically meaningful statement. The correlations
are invariant under reparameterizations of time. These correlations
can be determined by de-parameterizing the classical equations
to remove the reference to $t$. For the model under consideration, 
we arrive at a differential
equation governing the evolution of the scalar field as a function
of the scale factor by noting $d \phi/d a = \dot{\phi}/{\dot{a}}$
which from the Friedmann equation gives
\begin{equation} \label{dphicl}
	\frac{d \phi}{da} = \pm \sqrt{\frac{6}{\kappa a^2}} \;.
\end{equation}
Integrating this we find that the scalar field evolves as
\begin{equation}\label{phicl}
	\phi_{\mathrm{cl}}(a) = \pm \sqrt{\frac{6}{\kappa}} \log(a) + C
\end{equation}
where $C$ is a constant. The time parameter $t$ showing up
in the classical equations of motion can now be seen
an arbitrary parameterization of the correlations
between the physical quantities given in (\ref{phicl}).

To recover a notion of dynamics without any explicit
reference to time in our model we can choose
the scale factor to play the role of a physical clock
since it is a monotonically increasing function as indicated
in equation (\ref{clmotion}). While on general grounds it is not necessary
to interpret the quantum theory by singling out a clock
variable, in our model it proves useful for interpreting
our results in terms of a ``time evolution" of the scalar field
given in equation (\ref{phicl}). Thus we can consider
wave packets peaked around a value of the scalar field
at a given scale factor and then the difference equation
of LQC will determine the trajectory of the wave packet
for different values of the scale factor from
which we can determine how the scalar field evolves
with respect to the scale factor.

With an understanding of the classical framework 
we can now turn to the loop quantization
of the model.
In the framework of Dirac quantization used in LQG and LQC
the physical quantum states are annihilated by
the Hamiltonian constraint (\ref{Hclass}) represented
as an operator. The gravitational side of the constraint
when quantized using loop techniques leads to a partial
difference equation \cite{Bohr}.
The matter term is quantized as
\begin{equation}
	\widehat{H}_{\phi} =  \half \widehat{a}^{-3} \widehat{\Pp}^2
	 = -\half d_J \,  \frac{\partial^2}{\partial \phi^2}
\end{equation}
with $d_J$ being the LQC quantized eigenvalues of
the inverse volume operator. The key feature of
the $d_J$  is that it is a bounded function
with maxima at a characteristic scale factor
$a_* = \sqrt{\frac{8 \pi \gamma J \mu_0}{3}}\,\lp$. 
The inverse volume eigenvalues are labeled with the
quantum ambiguity parameter $J$ which arises from the fact
that the inverse volume operator is computed by tracing over
SU(2) holonomies in an irreducible spin $J$ representation \cite{dj}.
Physically, below the scale factor $a_*$ determined by $J$, the
inverse volume eigenvalues are suppressed in contrast to the
classical inverse volume which diverges for small $a$. 
$\mu_0$ is an additional quantum ambiguity parameter which 
is heuristically related to the smallest eigenvalue of area operator in LQG \cite{Bohr} (which fixes its value to $\sqrt{3}/4$). The eigenvalues  $d_J$ can be approximated as
$d_{J}(a)=  D(a)\, a^{-3}$ \cite{dj} with
\begin{eqnarray}
&&D(a) = \left( {8/ 77}\right)^6 q^{3/2} \Big\{7 \Big[(q+1)^{11/4}
-|q-1|^{11/4}\Big] \nonumber \\
&&~~~{}- 11q\Big[(q+1)^{7/4}-{\rm sgn}\,(q-1) |q-1|^{7/4}\Big]
\Big\}^6\!,\label{D}
\end{eqnarray}
where $q\equiv {a^2 /a_*^2}$. For $a>a_*$ we have $D(a) \approx 1$
and we recover the classical expression for the inverse volume
eigenvalues $d_J(a) \approx a^{-3}$.

The constraint equation satisfied by the physical
states $\psi_{\mu}(\phi)$ becomes 
\begin{widetext}
\begin{eqnarray}\label{diffeqn}
	\frac{3}{4 \kappa \gamma^2 \mu_0^2} \Big[
	s(\mu+4 \mu_0) \,\psi_{\mu+4 \mu_0} - 
	2 s(\mu) \, \psi_{\mu} + s(\mu-4\mu_0) \, \psi_{\mu-4\mu_0} \Big]
	 - \frac{1}{2} d_J(\mu) \frac{\partial^2 \psi_{\mu}}{\partial \phi^2} = 0
\end{eqnarray}
\end{widetext}
where the parameter $\mu$ is an eigenvalue of the scale factor
operator  and volume operator given by
\begin{equation}
	a = \sqrt{\frac{8 \pi \gamma \,|\mu|}{6}} \lp \,, ~~
	V_{\mu} = \left(\frac{8 \pi \gamma |\mu|}{6} \right)^{3/2}
	\lp^{\,3} \label{a_mu}
\end{equation}
and the function 
$s(\mu) = \frac{2}{8 \pi \gamma \mu_0 \lp^2}(V_{\mu+\mu_0}-V_{\mu-\mu_0})$
can be shown to be equal to $\sqrt{p}=a$ for large volumes \cite{kevin_ham}.
We note that this difference equation is derived from a non-self adjoint 
constraint operator. One can construct a self-adjoint constraint \cite{Willis, time} and derive the corresponding difference equation, however we do not expect our results to be modified significantly. 
The difference equation can also be seen as a discrete approximation
of the second order hyperbolic Wheeler-DeWitt equation
\begin{equation}\label{WDeqn}
	\frac{\kappa}{3} \frac{\partial^2}{\partial p^2} 
	\Big( \sqrt{p} \, \psi(p,\phi)\Big) - 
	\frac{1}{2} d_J \frac{\partial^2 \psi(p,\phi)}{\partial \phi^2}
\end{equation}
obtained by quantizing $\hat{c} = i  \frac{1}{3} \gamma \kappa 
\, \partial/\partial p$. It has been shown in Ref. \cite{Kiefer}
that wave packets which satisfy the Wheeler-DeWitt equation 
(though for a different factor ordering than that given here and
without $d_J$ corrections)
follow the classical trajectory given in equation (\ref{phicl}).
The constraint equation (\ref{diffeqn}) differs radically
from the Wheeler-DeWitt equation by the fact that it is a
discrete difference equation and the presence of
the inverse volume eigenvalues $d_J$. It is our goal now
to determine the precise effects of these differences in
the trajectory of the wave packets.

Attempts have been made to describe the new features
of the quantum theory of LQC within an effective continuous
theory. The very first attempts simply replaced
the inverse volume in the classical equations of motion
with the eigenvalues $d_J$ \cite{superinflation}. Many of
the phenomenological investigations are based on this
effective framework. The effective equations have been
generalized both through a path integral framework \cite{kevin_ham}
and a WKB analysis \cite{WKB,WKB2} to include effects arising
from the fundamental discreteness of the theory. The effective framework can
be described in terms of an effective Hamiltonian constraint
given by (additional terms arise in WKB analysis, which are 
not considered here)
\begin{equation} \label{Heff}
        H_{\mathrm{eff}} = - \frac{3}{\kappa \gamma^2 \mu_0^2}  \sqrt{p} \; \sin^2(\mu_0 c)
        + \half d_J(a) \Pp^2
\end{equation}
where the $\sin^2(\mu_0 c)$ modifications can be understood
as discreteness corrections from the difference equation.
Most phenomenological investigations so far have
ignored the discreteness corrections 
by assuming $\mu_0 c \ll 1$ and $\sin(\mu_0 c) \approx \mu_0 c$
whence the modified Friedmann equation becomes
\be
\left(\frac{\dot a}{a}\right)^2 = \frac{\kappa}{3} \, \rho_m = \frac{\kappa}{3} \, \frac{\dot \phi_{{\mathrm{eff}}}^2}{2 D(a)} \label{modH}
\ee
where $\rho_m = \half d_J \Pp^2 / a^3 = \half D(a) \Pp^2$
and we have used $\dot \phi_{\mathrm{eff}} = d_J P_\phi$ acquired
from the Hamiltonian equations. It is then straightforward to
deparameterize the equations of motion to obtain
\begin{equation}
	\frac{d \phi_{{\mathrm{eff}}}}{da} = \pm \sqrt{\frac{6 a \, d_J(a)}{\kappa}} \; \label{et1}.
\end{equation}
For $a < a_*$, $d_J(a)$ is proportional to positive powers of the scale factor which leads to radical modifications
of the  $\phieff(a)$ trajectory compared to the one obtained classically.
For $a$ larger than $a_*$, we have $d_J(a) \approx a^{-3}$
and we recover the classical behavior of the scalar field given in 
eq. (\ref{phicl}).

In our analysis we would also like to consider the discreteness corrections from
the gravitational side of the constraint.
From the vanishing of the effective constraint (\ref{Heff})
it is clear that the discreteness corrections become
relevant when the matter term becomes large. More
precisely when the matter density $\rho_m$ is on
the order of a critical density $\rho_{\mathrm{crit}} 
\equiv 3/\kappa\mu_0^2\gamma^2 a^2$, the corrections
are appreciable \cite{kevin_ham}.
The equations of motion can be calculated from
the effective constraint (\ref{Heff}) and 
deparameterized to give
\begin{eqnarray}
	\frac{d \phieff}{d a} &=& \pm \sqrt{\frac{6 a d_J}{\kappa} \frac{1}{\left(1 - \frac{\kappa \gamma^2 \mu_0^2 d_J P_\phi^2}{6 a}\right)}}  \nonumber \\
	&=& \pm \sqrt{\frac{6 a d_J}{\kappa} 
	\frac{1}{\left(1- \rho_m / \rho_{\mathrm{crit}}\right)}} \label{et2}
\end{eqnarray}
whence it is clear that the modified equations of motion
in equation (\ref{et1}) are recovered when $\rho_m \ll \rho_{\mathrm{crit}}$.
The Friedmann equation with discrete quantum corrections can be
obtained from the above equation and is given by \cite{kevin_ham}
\be
\left(\frac{\dot a}{a}\right)^2 = \frac{\kappa}{3} \, \rho_m - \frac{1}{9} \kappa^2 \gamma^2 \mu_0^2 a^2 \rho_m^2 ~.
\ee
For simplicity we will refer to effective theories described by eq.(\ref{et1}) and eq.(\ref{et2}) as ET-I and ET-II
respectively.

As emphasized earlier, the use of these effective continuous equations
requires a more careful consideration. Most phenomenological
investigations have so far assumed that the effective continuous equations
remain valid even  near the Planck regime, in particular
till the fundamental step size
of the difference equation $4 \mu_0$ which corresponds
a scale factor of $a_0 \approx \sqrt{16 \pi \gamma \mu_0 / 3} \, \lp$.
Therefore it is crucial to determine under what conditions
do the effective equations hold and the 
phenomenological predictions can be trusted.
It is important to  determine the scale at which discrete 
quantum geometric effects 
play a prominent role and influence dynamics. We can then answer the question 
as to what scale does the continuum spacetime arise and classicality emerges. 
We can also determine if further corrections arise from pure
quantum effects. 

\section{Coherent state evolution and quantum dynamics}

Our aim is to
compare the classical and effective theories with the evolution from
the difference equation, and thus 
for simplicity we do not consider the evolution beyond the classical 
singularity. The difference equation (\ref{diffeqn}) is
sufficiently complicated
such that an analytic solution is not available. We will thus
compute the solutions numerically. 
In our method we consider a semi-classical state at a large initial
scale factor and evolve it backward toward the singularity
using the difference equation.
As a semi-classical state we consider a Gaussian wave packet
sharply peaked around $\phi = 0$ and some classical $\Pp$ at scale factor $a_{\mathrm{init}} \gg a_*$
\begin{equation}
	\psi_{a_i}(\phi) = \exp(i \Pp \phi) \exp(-\phi^2/ 2 \sigma^2)
\end{equation}
with a spread $\Delta \phi = \sigma$. Since the difference equation is second order in $\mu$, to
find a physical solution
we must specify initial conditions at $\mu_{\mathrm{init}}$ (determined from $a_{\mathrm{init}}$ using eq.(\ref{a_mu})) 
and $\mu_{\mathrm{init}} - 4 \mu_0$. The difference
equation then gives us the wavefunction at $\mu_{\mathrm{init}} - 8\mu_0$ which serves to determine wavefunction
at the next step and so on, thus yielding us evolution of the initial Gaussian. 
The scalar field trajectory $\phi(a)$ will then be obtained
from the peak of the semi-classical state (we will comment on the validity
of this in the discussion).

Given the Gaussian initial condition, exact solutions can be computed
numerically by assuming the form 
\begin{equation}
	\psi_{\mu}(\phi) = \exp(i \Pp \phi) \exp(-\phi^2/ 2 \sigma^2)
	\sum_{n=0}^{\infty} C_{n}(\mu) \phi^n ~.
\end{equation}
With the above ansatz, the  partial difference equation is reduced to a 
difference
equation for the coefficients $C_n (\mu)$ which can be solved
numerically (the ansatz avoids computation of  finite differences of 
$\partial^2 \psi_{\mu}/  \partial \phi^2$).
The initial condition is then simply $C_n(\mu_{\mathrm{init}}) = \delta_{n 0}$
for some large  $\mu_{\mathrm{init}}$. 
We are left to specify the wavefunction at $\mu_{\mathrm{init}}-4\mu_0$.
This choice does not affect our results appreciably.
We specify the initial condition by analogy with the Klein-Gordon
equation where an arbitrary solution is the sum
of positive and negative frequency solutions (which
correspond to the $\pm$ solutions of equations
(\ref{phicl},\ref{et1},\ref{et2})).
We can then tune the initial conditions to pick
out one of the two solutions.

\begin{figure}[ht]
\begin{center}
\includegraphics[width=8cm, keepaspectratio]
        {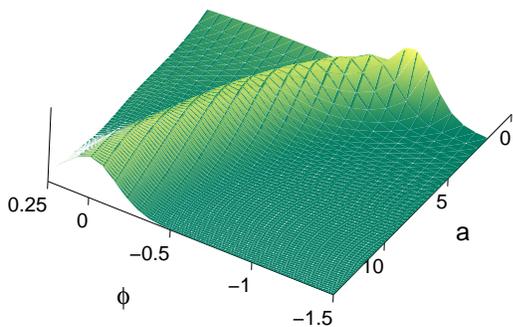}
\end{center}
\caption{Evolution of coherent state for $J=1/2$ and $P_\phi = 10 l_p$. The coherent state remains sharply peaked and follows the classical
trajectory given by equation (\ref{phicl}). Amplification at scales close to $a = 0$ are due to discrete quantum effects.}
\label{fig1}
\end{figure}

\begin{figure}[ht]
\begin{center}
\includegraphics[width=8cm, keepaspectratio]
        {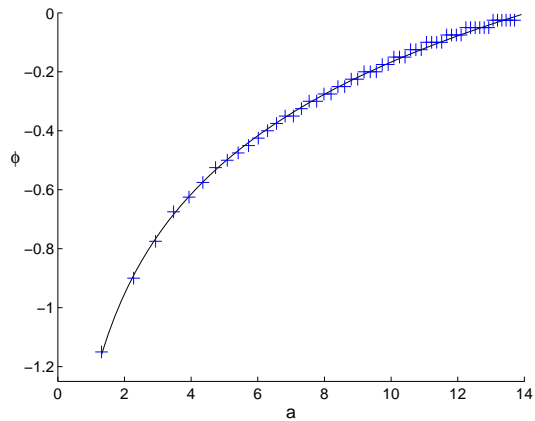}
\end{center}
\caption{The trajectory of the peak of coherent state ($+$ curve) is compared with classical theory
(solid curve) for $J = 1/2$ and $\Pp = 10 l_p$. The classical trajectory agrees extremely well with the quantum curve until the smallest discrete 
step in the scale factor.
}
\label{fig2}

\end{figure}

For the given initial coherent state peaked at $\phi = 0$ and $P_\phi$ we can also evaluate 
the classical and effective trajectories from ET-I and ET-II and compare them with the trajectory of the
peak of the coherent state. 
For larger values of $J$, $a_*$  increases and we thus expect
to see deviations from the classical dynamics at larger scales (yet still below $a_*$).
The deviations from ET-II are expected when the matter
density approaches the critical value. We can test these corrections
by choosing a large initial value of $\Pp$.
Furthermore, since the fundamental discrete step is of size $4 \mu_0$ we do not
expect classical or effective theories to be valid for $\mu \lesssim 4 \mu_0$. 
We now discuss some representative cases from our numerics:

(i) $J=1/2, P_\phi = 10 l_p$: This case corresponds to the smallest value of $J$ and a small value of $P_\phi$ with
$\mu_{\mathrm{init}} = 200$. The matter density remains small compared
to the critical value which ensures that differences between ET-I and ET-II are negligible throughout the evolution. Since $J=1/2$, we have  $a_* < a_0$ which
implies that the effective dynamics agrees with the classical dynamics for all scale factors.
The evolution of the coherent state via the quantum difference equation in 
shown in Fig.\ref{fig1}. 
The coherent state is sharply peaked at $\mu_{\mathrm{init}}$ and evolves toward $\mu = 0$ without loosing its semi-classical
character and retaining its sharp peak. The trajectory of the peak  is compared to the 
classical and effective theories  in Fig.\ref{fig2}. 
 The classical and effective theories are
in very good agreement with quantum theory till the smallest non-zero value of $\mu$.
It is clear that for this choice of
parameters, the classical evolution can be trusted till the first step in 
the quantum evolution before the classical singularity,
i.e. till $a_0$. Below $a_0$ the classical evolution would lead to a 
blow up of $\phi$  resulting in a singularity 
whereas the evolution is non-singular with the quantum difference
equation. For the chosen value of parameters, 
classicality and continuum thus emerge as soon as we consider a scale 
factor greater than $a_0$.

\begin{figure}[ht]
\begin{center}
\includegraphics[width=8cm, keepaspectratio]
        {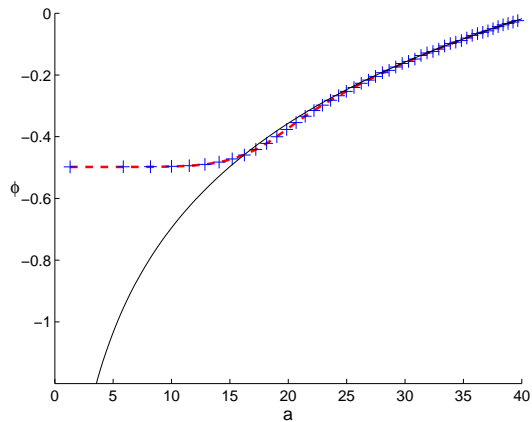}
\end{center}
\caption{Comparison of evolution via quantum constraint ($+$ curve), ET-I and ET-II (dashed curve)
and classical theory (solid curve) for $J = 500, \Pp = 100 l_p$. Classical theory breaks down for $a < a_*$, whereas ET-I and ET-II
are valid till $a_0$.
}
\label{fig3}

\end{figure}

(ii) $J = 500$, $\Pp = 100 l_p$: Here we start the evolution  at $\mu_{\mathrm{init}} = 350$ 
which corresponds to an initial scale factor twice $a_*$.
In this case $a_* > a_0$ and we expect a region where the classical theory 
breaks down and dynamics can be
approximated by the effective theory. We have chosen $\Pp$ and $\mu_i$ in such a way that differences between
ET-I and ET-II are negligible even for very small $\mu$. 
The results are plotted in Fig.\ref{fig3}. 
It is evident that the classical theory  departs from the quantum 
evolution at a larger scale factor as compared to the case of $J =
1/2$. 
The scale at which departure becomes significant 
is $a_*$. It should be noted that ET-I and ET-II  agree with the 
quantum difference
evolution till the smallest step size in scale factor. Thus, for 
large $J$ and reasonably chosen $\Pp$ such that
$\rho_m \ll \rho_{\mathrm{crit}}$ during evolution, ET-I and ET-II 
can be trusted to the smallest allowed scale factor $a_0$.
Unlike case(i), classicality emerges for scale factors $a > a_*$, 
however continuum spacetime is a good approximation
to discrete quantum geometry for $a > a_0$.

\begin{figure}[ht]
\begin{center}
\includegraphics[width=8cm, keepaspectratio]
        {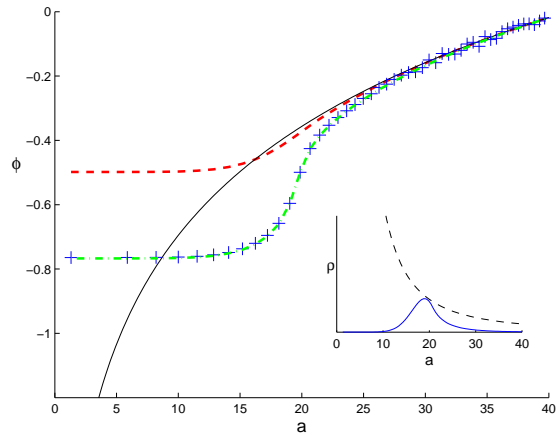}
\end{center}
\caption{Evolution for $J=500$, $P_\phi = 1600 l_p$. The classical (solid curve) and ET-I trajectory (dashed curve) 
deviate from
quantum evolution ($+$ curve) for scale factors at which perturbative corrections become significant. ET-II agrees
with evolution from quantum constraint till $a_0$. Inset: 
ET-II corrections become significant when the matter
density (solid curve) approaches the critical density (dashed curve).
}
\label{fig4}

\end{figure}

(iii) $J=500$, $\Pp = 1600 l_p$: As in case(ii) we start the evolution from $\mu_{\mathrm{init}} = 350$. However, we now
choose $\Pp$ such that the matter density becomes on the
order of the critical value for scale factors 
greater than $a_*$. We thus expect significant departure between the 
quantum evolution from 
classical theory and ET-I for larger scale factors compared to those 
in case (ii). 
The results are shown in Fig.\ref{fig4}. The ET-II corrections to the 
Friedmann dynamics 
become significant at $a \sim 20 l_p$ where both the classical 
 and ET-I dynamics
start to disagree with the  quantum evolution. 
ET-II  which incorporates these corrections matches 
very well the coherent state evolution till $a_0$. The inset of the 
 figure shows the evolution of the matter density
with the scale factor. 
The dashed curve represents the critical density $\rho_{\mathrm{crit}}$ and
it is clear that near $a \sim 20 l_p$, the matter density becomes
on the order of the critical value and precisely there the deviations
due to ET-II arise.

The picture which emerges from these cases is confirmed in various 
other numerical studies we performed. All of 
the numerical results can be broadly classified in the above 
three cases. In summary our results show: \\
(i) The classical evolution matches very well with the quantum
evolution 
till very small scale factors. For
small $J$ and small $P_\phi$, the classical theory can even be trusted 
till $a_0$. For large $J$ and/or large
$\Pp$, departures occur at larger scale factors. \\
(ii) The effective theories with appropriate corrections are very 
good approximations to difference
equation till scale $a_0$. With suitable choice of initial 
conditions such that the gravitational corrections of ET-II 
are negligible, various phenomenological applications based on ET-I 
are trustworthy. \\
(iii) Since effective theories are good approximations for $a > a_0$, 
the assumption that continuum spacetime 
emerges above $a_0$ proves  reasonable. Emergence of classicality, 
however depends on various factors. For small
$J$ it may occur for $a \sim a_0$ (if $\Pp$ is not large). For large 
$J$ it emerges at larger values of the scale factor.

\section{Discussion}
Let us summarize the main results presented in this work. We were
able successfully to construct semi-classical states within
the discrete framework of LQC which were represented as
localized wave packets that for large volumes followed
the classical trajectory. This can be viewed as
a systematic  demonstration that LQC has
the correct semi-classical limit while leading
to modifications in the high curvature regime of the 
universe. Our results indicate that the discrete
nature of the universe predicted by LQC can be adequately
described in terms of an effective continuous picture
down to the Plank scale. Since the quantum theory of LQC
predicts discrete steps of the order of the Planck length,
it can safely be said that the continuous picture breaks down below
the Planck scale. The surprising
conclusion that follows from our results is that the continuous 
emergence occurs very near the 
Planck scale. This does not imply however that
above the Planck scale we recover the standard Friedmann
dynamics. The emergence of the classical spacetime
depends on choice of $\Pp$ and $J$ in initial configuration and 
following our procedure the scale of classical emergence can be
found in a straightforward way.

We have also demarcated explicitly the domain of validity of 
effective descriptions ET-I and
ET-II. The former was based on including modifications to the inverse scale factor 
to the standard Friedmann dynamics, whereas the latter also includes the
modifications pertaining to large extrinsic curvature. For the choice of
initial conditions such that modifications due to extrinsic curvature
are negligible, ET-I serves as a good effective theory. Most of the
phenomenological applications investigated so far in LQC fall into
this category. However, for generic initial conditions ET-II is a more
faithful description of the quantum theory of LQC. This is the new
phenomenological input in LQC and opens a new avenue for various applications.

Let us compare our results with earlier proposals for effective dynamics.
In one approach  kinematical coherent states
are constructed and then an effective Hamiltonian constraint
is calculated from the expectation value of the constraint
operator acting on these states \cite{Willis}.  The relation between the 
kinematical coherent states in that work and the coherent states
considered in this paper is not clear since the kinematical coherent
states are not annihilated by the Hamiltonian
constraint operator and are thus not physical states.
There is some evidence that a technique known as group
averaging to ``project"
kinematical states into physical ones (and in the process
provide a physical inner product which we discuss later), 
will project the kinematical coherent states into physical
ones with similar properties \cite{groupavg}, thus
providing a connection between their work and ours. 
Further, in the kinematical coherent state approach additional
corrections are calculated which depend on the spread of the
coherent state. Effects of these corrections have not been
considered in this paper. Interestingly, 
the obtained modifications to the classical equations are very similar in 
both approaches and a more detailed analysis is needed to make
a complete comparison.

A second approach has been to consider a WKB approximation
for solutions to the difference equation and from that
extract an effective Hamiltonian constraint \cite{WKB,WKB2}.
Through this technique one calculates the same
effective Hamiltonian constraint as that given
in eq. \ref{Heff} plus an additional term
denoted by the quantum geometry potential. 
Our results of testing
ET-II are directly applicable to this approach and suggest that
phenomenological applications based on WKB analysis occurring at
scale factors greater than $a_0$ due to inclusion of higher order 
terms in $\sin(\mu_0 c)$ expansion (as in eq.(\ref{Heff})) seem
trustworthy. Since, the evolution occurs with discrete steps of the order
of $a_0$, any effect based on continuum dynamics below this scale requires
further justification. 
In this work we have not tested the effects of
additional potential in WKB Hamiltonian constraint.
A careful and detailed analysis is required to understand
the significance of the derived potential in this approach.

Finally another method has been to consider the evolution
of kinematical coherent states by evolving them
with the unitary operator acquired by exponentiating
the Hamiltonian constraint operator \cite{time}. The effect is to
introduce a coordinate time parameter into the quantum theory.
This is equivalent to gauge fixing the lapse to unity
and then quantizing the remaining Hamiltonian
as an {\em unconstrained} system which leads to a difference
equation with a Schrodinger equation like $\partial/\partial t$
term on the right hand side. Performing this procedure, dynamics
is easy to describe in terms of the coordinate time parameter
now appearing in the difference equation. However, the 
framework 
presented in this paper has the advantage
that the wavefunctions are physical states, i.e. those that
satisfy the difference equation. This is a crucial feature
which testifies to the validity of our results. Within
the method considered here, various continuous effective
equations of motion can be tested in a precise manner.
Since explicit gauge fixing
is involved in analysis of Ref. \cite{time}, a comparison of those results 
 with the ones obtained  here is difficult.

Let us turn to an open issue in the program presented here. In
extracting the trajectory of the scalar field from the
wavefunctions we have used the peak of the wave packet. Properly
done, the trajectory should be calculated as an expectation
value of the operator $\hat{\phi}$ with a suitable
probability measure provided by the physical inner product. 
A procedure for providing a physical inner product
that is suitable for constrained systems like
LQC is known as group averaging \cite{GA} and this
technique has been explored in depth in a model of
LQC with a cosmological constant in \cite{npv_model}.
We can  speculate as to the effect of using the correct
probability measure by noting the similarity of the difference
equation to the Klein-Gordon equation. The application
of group averaging to the Klein-Gordon constraint
yields a probability measure that is time independent
and {\em positive definite} on the space of both positive and
negative frequency solutions. It can be
shown that when considering coherent states
which are sharply peaked, the expectation value
of the scalar field using the group averaging
probability measure is approximately equal to the peak
of the wave packet which is evidence that we
do not expect the results of this paper to be significantly
modified by using the correct probability measure. This
does not rule out the possibility of corrections arising from
the measure in the small volume regime where discreteness
effects may play an important role in the physical inner product.

This work opens the possibility for further quantum applications of semi-classical
nature. Apart from previously considered phenomenological applications, it will
be interesting to consider for instance anisotropic models where one of the
geometrical degrees of freedom can serve as a clock and semi-classical wave-functions
can be constructed similarly. 
Furthermore the dynamics of ET-II suggests the existence of bouncing and recollapsing phases when the matter density 
becomes large. This also opens
the possibility for the presence of tunneling regions the description of which would require the
understanding of the quantum evolution of semi-classical states.
Additional applications include investigations of the evolution
through the singularity and the relation to pre big bang scenarios.

\acknowledgements{We would like to thank Abhay Ashtekar for useful discussions 
and helpful comments, and Martin Bojowald for a careful reading
of the
manuscript and helpful comments. This work was supported by the Eberly research funds of Penn
State and NSF grants PHY-0090091 and PHY-0354932.}

\end{document}